# CONSTRUCTION OF QUANTUM WAVEFUNCTIONS FOR NON-SEPARABLE BUT INTEGRABLE TWO-DIMENSIONAL HAMILTONIAN SYSTEMS BY MEANS OF THE BOUNDARY VALUES ON THE CLASSICAL CAUSTICS

M. Fusco Girard

*Department of Physics "E. R. Caianiello",*

*University of Salerno*

*and*

*Gruppo Collegato INFN di Salerno,*

*Via Giovanni Paolo II,*

*84084 Fisciano (SA) ITALY*

*Electronic address: mario.fuscogirard@sa.infn.it*

## ABSTRACT

It is shown that it is possible to construct the quantum wave functions for non-separable but integrable two-dimensional Hamiltonian systems, by solving suitable Dirichlet boundary values problems inside and outside the regions spanned by particular families of classical trajectories, in one-to-one correspondence with the quantum state. The method is applied both to the Schrödinger equation, and to the quantum Hamilton-Jacobi equation. The boundary values are obtained by integrating the 1-dim equations on the caustics' arcs enveloping the classical trajectories. This approach gives the same results as the usual methods, and furthermore clarifies the links between quantum and classical mechanics.

## 1. INTRODUCTION

In a recent paper [1], it was shown that it is possible to compute the energy eigenvalues for a non-separable but integrable two-dimensional Hamiltonian system, by solving the 1-dimensional Schrödinger Equation (SE), or the equivalent Quantum Hamilton-Jacobi Equation (QHJE), on the caustics' arcs enveloping particular families of classical trajectories with that energy. Both these approaches are rigorously in the framework of the usual Copenhagen interpretation of the quantum mechanics. The values so obtained for the energy levels are the same as given by traditional methods. The aim of the present paper is to show that this procedure can be extended to construct the quantum wave functions too, by solving suitable Dirichlet boundary problems for the SE or for the QHJE.

The starting point is the assumption that to every quantum wave function for a time independent integrable Hamiltonian system it corresponds, in the configuration space, a particular family of classical trajectories, enveloped by a caustic. The link between them is given by the complex phase $W_Q(\mathbf{q}, E)$ of the wave function. Here $\mathbf{q}$ denotes the set of the n spatial coordinates and E is the energy. $W_Q(\mathbf{q}, E)$ is the quantum Hamilton's Characteristic Function, also called the abbreviated action [2]. In the following, for simplicity it will be named the quantum action, and is a solution of the QHJE, which is equivalent to the SE. When the Planck's constant $\hbar$ goes to 0, the quantum action becomes a solution $W_C(\mathbf{q}, E)$ of the Classical Hamilton-Jacobi Equation (CHJE), i.e. the corresponding classical action; this in turn generates a family of classical trajectories, enveloped by a caustic C. In this way to each quantum state for these systems, it corresponds a particular caustic. These considerations hold for each number of degrees of freedom, and can be inverted. A typical two-dimensional example is reported in Fig. 1, which refers to a trajectories' family for the Barbanis Hamiltonian, with coordinates $\mathbf{q} = (x, y)$ and unitary mass:

$$H(x, y) = \frac{1}{2}(p_x^2 + p_y^2) + \frac{1}{2}(\omega_x^2 x^2 + \omega_y^2 y^2) + \lambda x^2 y \qquad (1).$$

The caustic C consists of four arcs, which meet the equipotential line $U(x, y) = E$ in four points, the caustic's vertices. The equipotential line encloses the region of the configuration space allowed to all the trajectories with the same energy E, while the caustic encloses the region $\Omega_F$ spanned by that particular family F of trajectories. For the case in figure, each point of $\Omega_F$ is crossed by two trajectories, so that the family F includes two sub-families. By orienting the trajectories of a sub-family, the momentum $\mathbf{p}$ becomes uniquely defined in every point of $\Omega_F$. This vector field $\mathbf{p}(x, y)$ is related to a special solution of the classical HJ equation, by the relation:

$$\mathbf{p} = \text{grad } W_C(x, y, E) \qquad (2).$$

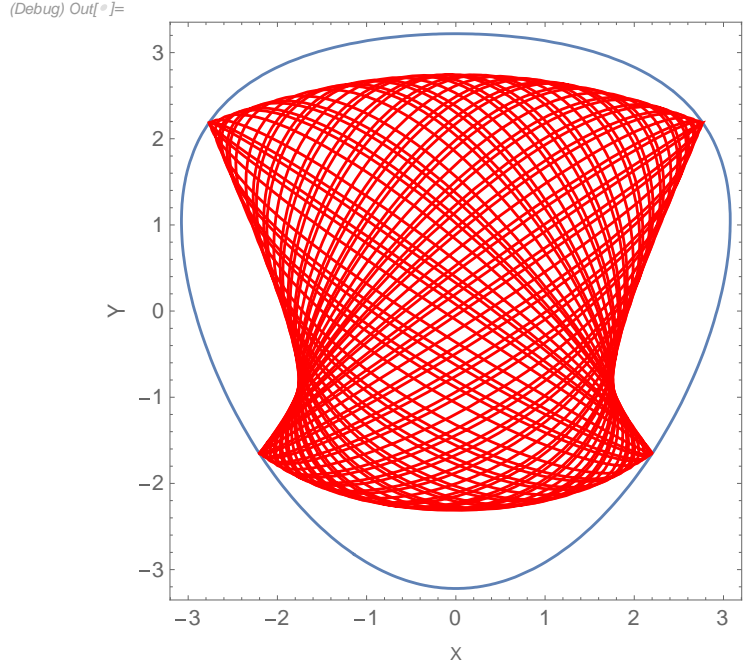

*Fig. 1. A family of trajectories for the Barbanis Hamiltonian (1). The parameters are $\omega_x = 1.1$, $\omega_y = 1.0$, $\lambda = -0.11$. The energy is E = 5.18266.*

In this way, to each oriented sub-family of trajectories it corresponds a classical action $W_C(\mathbf{q}, E)$, uniquely defined apart from an unessential additive constant. $W_C(\mathbf{q}, E)$ is the limit for $\hbar \to 0$ of a unique quantum action $W_Q(\mathbf{q}, E)$, which in turn is the complex phase of a solution $\psi(\mathbf{q}, E)$ of the Schrödinger equation. In the generic case, this solution is not normalizable, but when the quantization conditions are satisfied, $\psi(\mathbf{q}, E)$ is a regular quantum wave function and in this way, the one-to-one link between a wave function and the corresponding family of classical trajectories is established. The trajectories' family connected to a wave function $\psi(\mathbf{q}, E)$ will be indicated as $F_\psi$, and $C_{F,\psi}$ will denote its caustic. Formally, analogous considerations hold externally to the caustic too, where $W_C(\mathbf{q}, E)$ and one or both the components of the momentum $\mathbf{p}$ are complex quantities.

In [1] it was shown that in the 2-dim case, with respect to the corresponding wave function, the caustic $C_{F,\psi}$ has a role analogous to the classical turning points in the 1-dim case. There, the second derivative of the wave function vanishes, and they separate the intervals of the real axis where the wave function has oscillating behavior, from those where it exponentially decreases. Analogously, the second derivative of the wave function, normal to the caustic $C_{F,\psi}$, vanishes on the caustic itself, and the caustic separates the region $\Omega_F$ of the configuration space where the wave function oscillates, from the external region, where it exponentially decreases (monotonically or oscillating). This is obviously true for a separable 2-dim system, but is confirmed by the numerical study of the wave functions for integrable but non-separable systems too. Due to the vanishing of the normal second derivative on the caustic $C_{F,\psi}$, we assume that along each caustic's arc the quantum motion can be treated as 1-dim, and the wave function on the arc can be found by integrating the one-dimensional SE or QHJE, with a potential energy that is the restriction of U(x, y) to that caustic's arc. In this way, it is possible to compute the energy eigenvalues, with the same results as the usual approaches [1]. In addition, both the equations give the values of the wave function on the caustic's arcs, which therefore can be used as boundary values for two Dirichlet problems for the 2-dim SE or QHJE, constructing in this way the wave functions in the plane, separately inside the region $\Omega_F$, and outside of it.

## 2. CONSTRUCTION OF THE WAVE FUNCTIONS IN THE PLANE BY MEANS OF THE SCHROEDINGER EQUATION

The procedure to construct a quantum wave function in the plane and the corresponding family of classical trajectories, starts by choosing a tentative value for the energy E, and a point $(x_V, y_V)$ on the equipotential line U(x, y) = E; this point has therefore **p** = 0 and in the generic case is a vertex of a caustic. Thereafter, the trajectory starting from it is computed by means of numerical integration of the Hamilton's equations. The caustic is obtained by recording the points where the determinant of two independent solutions of the Jacobi equation vanishes [3]. As known, the Jacobi equation gives the time evolution of the vector connecting two neighboring trajectories [4]. Each caustic's arc, labelled with an index k, it is then fitted by means of a smooth function y = $f_k$(x) or x = $f_k$(y) (in the latter case it is only necessary to exchange x and y in the equations below). The 1-dim time independent SE for the wave function $\psi_k(x)$ on the arc is (apices indicate derivatives with respect to the argument):

$$-\left(\psi_k''(x) - \frac{g_k'(x)}{g_k(x)}\psi_k'(x)\right) = \frac{2m}{\hbar^2}(E - U_k(x))g_k^2(x)\psi_k(x)$$

(3),

where $g_k$(x) is the scale factor:

$$g_k(x) = \sqrt{1 + f_k'^2(x)}$$

(4),

and $U_k$(x) is the restriction of the potential energy U(x, y) to the arc k:

$$U_k(x) = U(x, f_k(x))$$

(5).

If the arc is a segment of the x-axis, $f_k'(x) = 0$, and the usual SE, without the scale factor, is recovered.

The Schrödinger equations (3) is then numerically integrated along each arc, with the same energy E, and with its own 1-dim potential energy $U_k(x)$, as given by Eq. (5). If E is an energy eigenvalue and the trajectories' family is precisely the one $F_\psi$ associated to the corresponding quantum state, four regular 1-dim wave functions $\psi_k(x)$, one along each caustic's arc and its continuation, are obtained [1]. In this case, the second derivative of the wave function vanishes at the arc's ends, so that the each arc contains the full oscillating part of its 1-dim wave function $\psi_k(x)$. Otherwise, the procedure is repeated with another initial point. If no choice of this latter brings to four regular wave functions, one has to try with a different value of E, and so on, until the energy and the four arc wave functions are found with the wanted accuracy. In the Fig. 2, two of these arc wave functions for the caustic in Fig. 1 are plotted. The energy is E = 5.18266, and the initial vertex has coordinates (-2.204, -1.650). The left graph is the wave function on the upper arc of the caustic (k = 2). At right, the wave function on the two symmetric transverse arcs (k = 1, 3). The wave function on the lower arc (k = 4) is similar to the one for the upper arc, but contained in a smaller interval. These arc wave functions belong to the (2, 2) state of the Barbanis Hamiltonian.

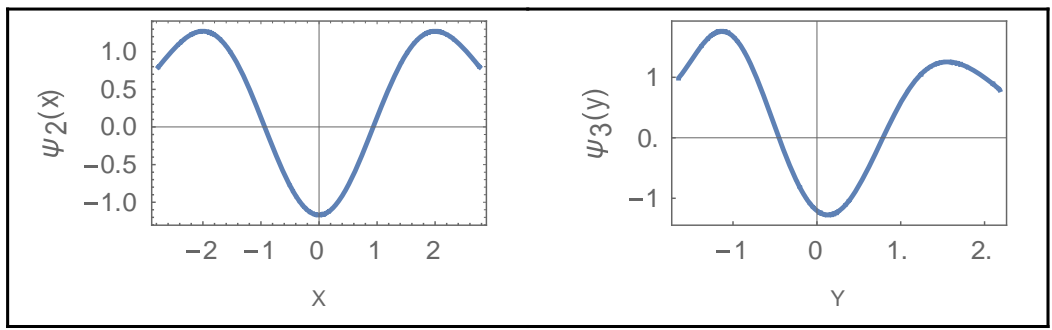

*Fig.2. Two of the arc's wave functions for the (2, 2) state of the Barbanis Hamiltonian. The arcs belong to caustic plotted in Fig. 1. The left graph is the wave function on the upper arc of the caustic. The right graph is the analogous for the two "vertical" arcs. Parameters' values as in Fig.1. The Planck's constant is ℏ =1.*

When the quantum energy and the caustic are found, the four arcs' wave functions $\psi_k(x)$ can be exploited as boundary values on the caustic, for two Dirichlet problems for the 2-dim SE, to construct the full wave function ψ(x, y) in the plane. The internal and the external Dirichlet problems, inside and outside the classical caustic, have to be separately considered, and the wave function for the entire plane is simply the union of the internal and the external solutions so found. Both the problems can be successfully solved, for instance, by means of a finite element method.

The procedure above has been successfully applied to many cases, and the figures [3-5] below present the results for the state (2, 2) with energy E = 5.18266 of the Barbanis Hamiltonian (1). The trajectories' family $F_\psi$ and the caustic $C_{F,\psi}$ are those plotted in Fig. 1. The boundary values employed on the caustic are the arc wave functions plotted in Fig. 2. Fig.3 reports the 3-dim plot of the internal wave function, inside the region enclosed by the classical caustic $C_{F,\psi}$, together with its contour plot. The parameters' values are the same as in Fig. 1.

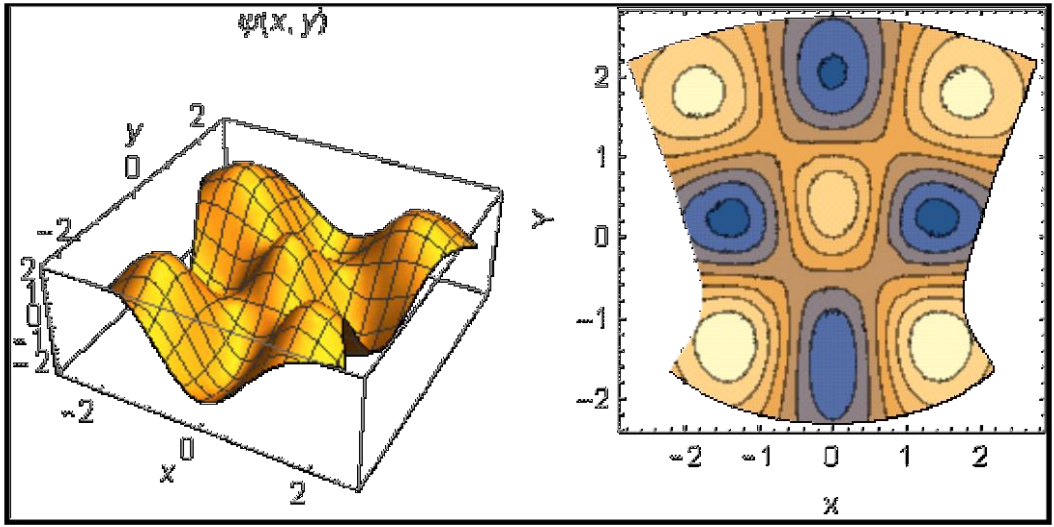

*Fig.3. The 3-dim plot of the internal wave function (2, 2) for the Barbanis Hamiltonian, inside the region enclosed by the relative classical caustic $C_{F,\psi}$, together with a contour plot of it. This wave function is the solution of the Dirichlet boundary problem in the region internal to the caustic, found as described in the text, by means of a finite element method. The boundary values are the arc wave functions plotted in Fig.2. The parameters' values are the same as in Fig.1 and $\hbar = 1$.*

The Fig. 4 analogously reports the solution of the external Dirichlet boundary problem, outside the caustic $C_{F,\psi}$, for the same state as in Fig. 3. The boundary values on the caustic are the same as for the internal problem, i.e. the arc wave functions in Fig. 2.

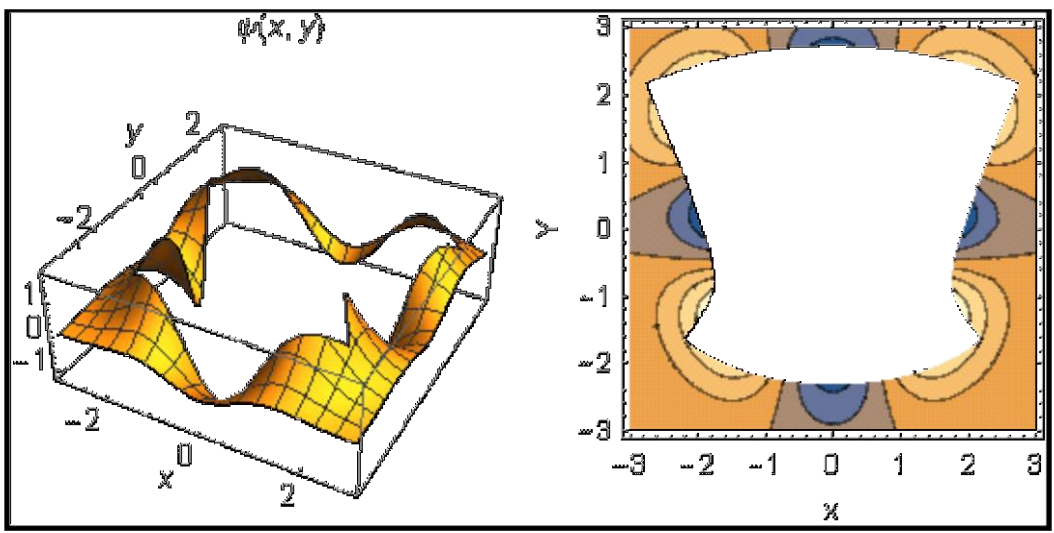

*Fig. 4. The 3-dim plot of the external wave function (2, 2) for the Barbanis Hamiltonian, outside the relative classical caustic $C_{F,\psi}$ (left), together with its contour plot. The parameters' values are the same as in Fig.1.*

The full wave function in the plane is obtained by the union of the internal and the external solutions. It is plotted in the Fig. 5 below, where the caustic $C_{F,\psi}$, along which the two functions are welded, is also shown.

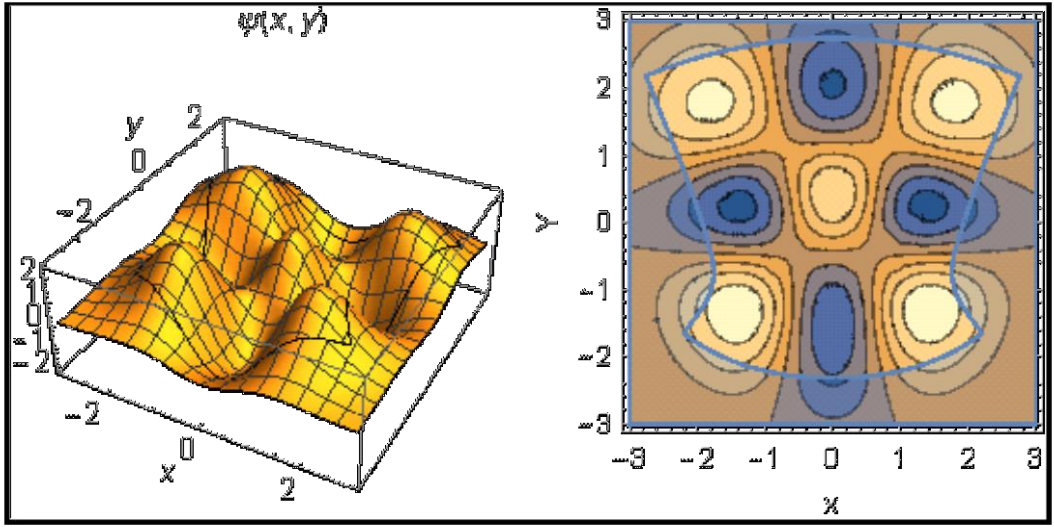

*Fig. 5. The full wave function (2, 2) for the Barbanis Hamiltonian, obtained by the union of the two partial solutions plotted in Figs. 3 and 4. The caustic $C_{F,\psi}$, where the internal and the external partial solutions are welded, can be seen in both the graphs.*

The energy value and the wave function obtained in this way, are the same as those obtained from diagonalization of the truncated Hamiltonian matrix. The same happens for all the wave functions computed by means of the above procedure. This agreement demonstrates the correctness of the initial assumption about the univocal correspondence between a quantum wave function and its family of classical trajectories.

3. CONSTRUCTION OF THE WAVE FUNCTIONS BY MEANS OF THE QUANTUM HAMILTON-JACOBI EQUATION

The analogous procedure to construct the wave function ψ(x, y) by means of the QHJE, is similar to the one for the SE, but a little more involved. The solutions of the SE in the plane are looked for in exponential form:

$$\psi(x, y) = \text{Exp}\left(\frac{i}{\hbar} W_Q(x, y)\right) \qquad (6).$$

The complex phase $W_Q(x, y)$, i.e. the quantum action, is written as:

$$W_Q(x, y) = X(x, y) + i\, Y(x, y) \qquad (7).$$

It is convenient to define:

$$A(x, y) = \mathrm{Exp}\left(-\frac{1}{\hbar} Y(x, y)\right) \qquad (8),$$

so that the wave function gets the familiar form:

$$\psi(x, y) = A(x, y)\ \mathrm{Exp}\left(\frac{i}{\hbar} X(x, y)\right) \qquad (9).$$

By inserting this into the time independent SE, and separating the real and imaginary parts, a system of two nonlinear partial differential equations follows [5]:

$$\frac{(\nabla X(x,y))^2}{2\,m} + (E - U(x, y)) = \frac{\hbar^2}{2\,m} \frac{\Delta A(x,y)}{A(x,y)} \qquad (10),$$

$$\nabla X(x, y) \cdot \nabla A(x, y) + \frac{A(x,y)}{2} \Delta A(x, y) = 0 \qquad (11).$$

Eq. (11) can be put in the form of a continuity equation. This system of equations has to be solved inside and outside the classical caustic, with the suitable boundary values for the involved quantities.

### 4. THE WKB APPROXIMATION

Firstly, by putting $\hbar = 0$, we get the equations for the WKB approximation of the wave function. In this case, the Eq. (10) becomes the classical HJ equation, and its solution $X_{cl}(x, y, E) \equiv W_C(x, y, E)$ is the classical action. The procedure is essentially that exposed in [3, 6], where the characteristic's curves method, i.e. the integration of the form **p.dq** along the trajectories, was exploited. As for the SE, in the present case too, one has to choose a tentative value for the energy and an initial vertex $v_1$ on the equipotential line, so to construct the trajectories' family and the caustic. Thereafter, the caustic' arcs are oriented from this vertex to the opposite one. The boundary values are obtained by applying the usual 1-dim WKB method on each caustic's arc. The classical momentum $p_{cl,k}(x, E)$ on the k-th arc is:

$$p_{cl,k}(x, E) = \sqrt{2\,m\,(E - U_k(x))} \qquad (12),$$

and is computed by recording the momentum of each trajectory touching the caustic. The corresponding classical action is:

$$X_{cl,k}(x, E) = \int p_{cl,k}(x, E)\, g_k(x)\, dx \; + const \qquad (13),$$

where the integration is done on the arc, from the initial point $x_{k,i}$ to the generic point $(x, f_k(x))$. The action is a monotonically increasing function along the arcs, from the initial vertex, where the action can be chosen equal to zero, to the opposite one, where it reaches its maximum value. Therefore, the constant is chosen equal to zero for the two arcs coming out from the initial vertex, and equal respectively to the final value of the action on the preceding arc, for the two other arcs. The classical actions have to satisfy the usual WKB conditions:

$$X_{cl,k}(x_{k,f}, E) - X_{cl,k}(x_{k,i}, E) = \int_{x_{k,i}}^{x_{k,f}} p_{cl,k}(x, E)\, g_k(x)\, dx = \hbar\,(n_k + 1/2) \qquad (14).$$

Here, $x_{k,i}$ and $x_{k,f}$ are the initial and final point of the k-th arc, and $n_k$ is an integer number. If the s.c. quantization conditions (14) are not satisfied, one has to try with another initial point, and possibly, to repeat the procedure with another tentative value for the energy and so on. In summary, an energy value E is semi classically allowed, if among all the trajectories' families with that energy, it there exists a particular one, whose caustics' arcs satisfy two relations (14), with two integers ($n_k$, $n_l$). As shown in [3, 6] these numbers are the same for each couple of opposite arcs. When the energy and the corresponding caustic are found with the required approximation, the classical actions $X_{cl,k}(x, E)$ on the caustics' arcs are exploited as boundary values for the equation (10) with $\hbar = 0$. The integration can be done, for instance, by means of a finite element method, to get the classical action $X_{cl,v1}(x, y, E)$ in the plane, both inside and outside the caustic. The second subscript remembers the initial vertex of the construction. The results of such a computation for the semi classical Barbanis state (2, 2) with the s.c. energy E = 5.18593 are shown in the Fig. 6. This action is obtained by orienting the caustic from the left lower vertex v1, to the opposite one. Thereafter, by putting $X_{cl,v1}(x, y, E)$ into the Eq. (11), the WKB amplitude $A_{WKB,v1}$ (x, y, E) of the wave function can be computed too. As happens in the 1-dim case at the turning points, this quantity diverges on the caustic.

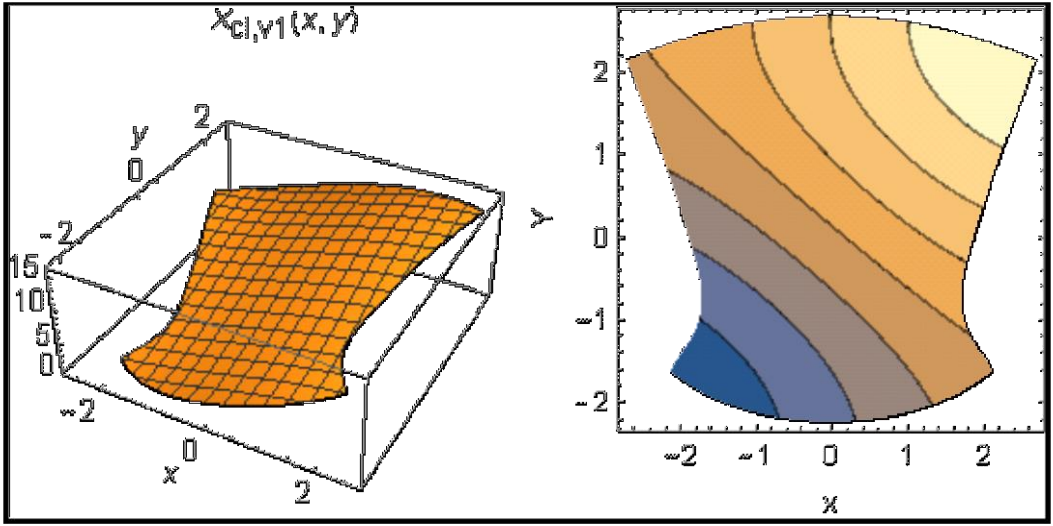

*Fig. 6. Left: the classical action $X_{cl,v1}$ (x, y, E) inside the caustic, for the trajectories corresponding to the WKB state (2, 2) of the Barbanis Hamiltonian, with the semi classical energy E= 5.18593. This action is obtained by orienting the caustic starting from the left lower vertex, indicated as v1. Right: the contour plot for the same function.*

In this way, the internal and the external Dirichlet problems for the Eqs. (10) and (11), with the Planck's constant put to zero, can be solved, so to construct a first WKB wave function. Inside the caustic, it is:

$$\psi_{WKB,v1}(x,y) = A_{WKB}(x,y)\, Sin(X_{cl,v1}(x,y)) \quad (15).$$

The amplitude $A_{WKB}(x,y)$ does not depend on the initial vertex. It is a slowly varying function well inside the classical region, and rapidly increasing near the caustic. On the caustic itself, it diverges. The inclusion of this amplitude in the expression (15) strongly deforms the wave functions for low-lying states. In order to avoid such effect, the figures below, which refer to the WKB (2, 2) state for the Barbanis system, are computed with a constant value of the amplitude. This first partial wave function is plotted in the Fig.7 below. For simplicity, only the internal function is shown.

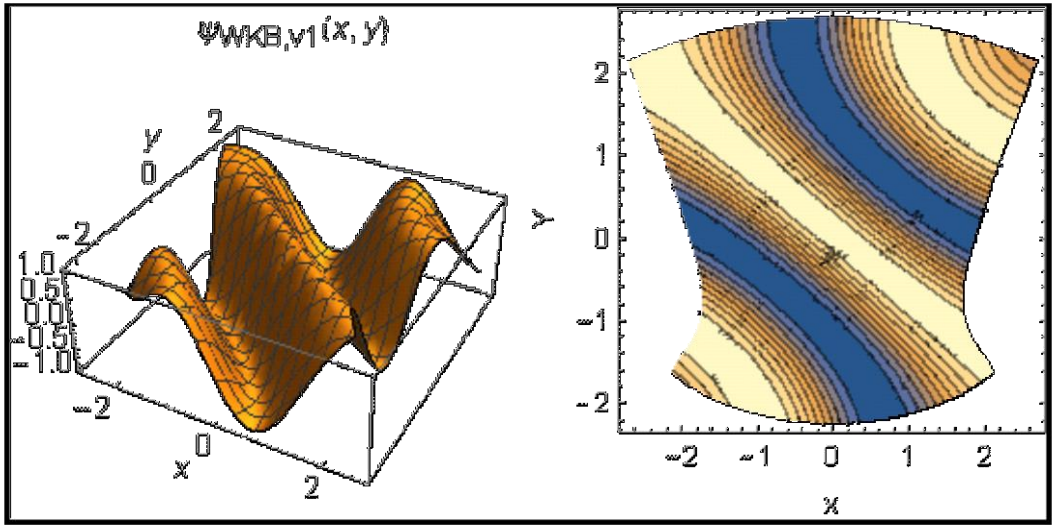

*Fig.7. The 3-dim plot and a contour plot for the first partial WKB wave function for the state (2, 2) of the Barbanis Hamiltonian, with the semi classical energy E= 5.18593. This function is obtained as explained in the text, by starting the construction from the left lower vertex v1 and with a constant value of the amplitude. Only the oscillating part, inside the caustic, is shown.*

For the Barbanis Hamiltonian, in order to get a wave function with the correct symmetry properties with respect to the x-coordinate, it is necessary to combine this first partial wave function with a second one, obtained by starting the construction with the caustic oriented this time from the x-symmetric vertex v2, which is the right lower one. The second classical action $X_{cl, v2}(x, y, E)$ is plotted in Fig. 8, while Fig. 9 reports the corresponding partial wave function.

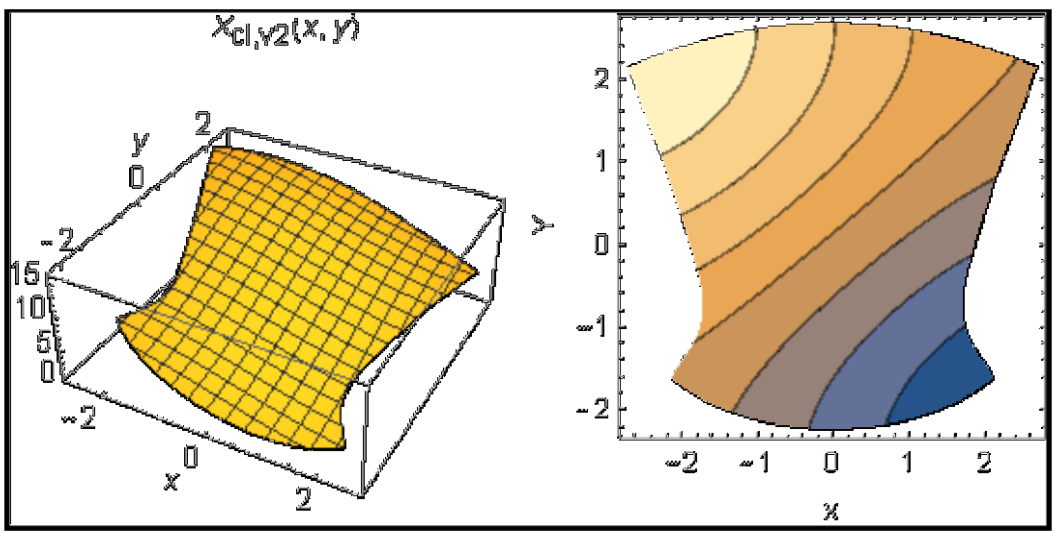

*Fig. 8. The same graphs as in Fig. 6, but referring to the second classical action $X_{cl,v2}(x, y)$ obtained with the caustic oriented starting from the right lower vertex, indicated as v2.*

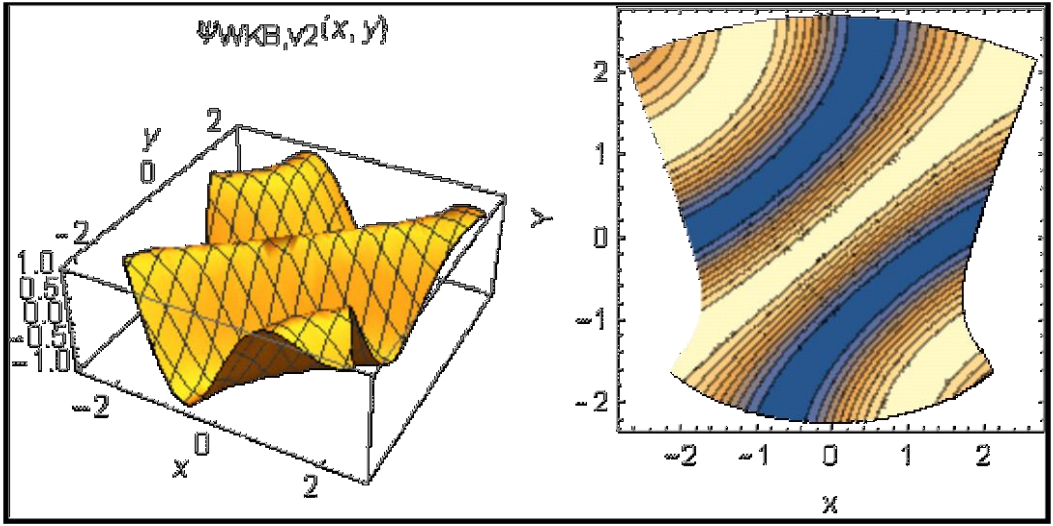

*Fig. 9. The second partial WKB wave function for the state (2, 2) of the Barbanis Hamiltonian, obtained by starting the construction from the right lower vertex v2, and with a constant value of the amplitude.*

The full x-symmetric or anti-symmetric WKB wave function is the combination:

$$\psi_{WKB}(x,y) = \psi_{WKB,v1}(x,y) \pm \psi_{WKB,v2}(x,y) \qquad (16),$$

where the upper sign is for x-symmetric wave functions and the lower one for the antisymmetric ones. This wave function, for the Barbanis semi-classical state (2, 2) computed by assigning a constant value to the amplitude A, is plotted in Fig. 10. The comparison between this figure and Fig. 3 shows that the WKB approximation preserves the structural properties of the true quantum wave function. The WKB energy computed with this method is $E_{WKB}$= 5.18593, to be compared with the quantum value E = 5.18266. The same wave function was computed in Ref. [3], by means of the characteristic curves method, instead of the finite element method employed here.

The two classical actions $X_{cl,v1}(x,y,E)$ and $X_{cl,v2}(x,y,E)$, employed to construct the two partial wave functions, are respectively linked to one of the two sub-families of trajectories which compose the family F. Both the sub-families enclosed in the same caustic so contribute to the full WKB wave function.

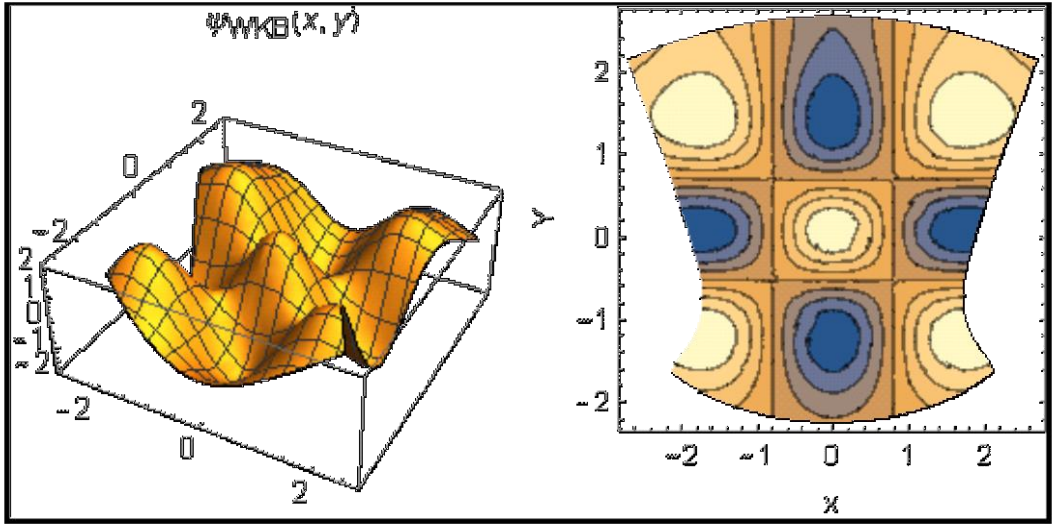

*Fig. 10. The full x-symmetric WKB wave function, obtained as explained in the text, by the sum of the two partial functions plotted in the Figures 7 and 9. Both the two sub-families of trajectories enclosed in the same caustic contribute to this wave function.*

## 5. THE CONSTRUCTION OF THE WAVE FUNCTIONS BY MEANS OF THE FULL QHJE

The construction of the quantum wave functions by means of the exact QHJE is similar to that exposed in the previous section, with the difference that the Planck's constant is maintained in Eqs. (10)-(11). The solution $W_Q(x, y, E)$ of these equations, i.e. the quantum action, is the complex phase of the wave function but, as discussed in [7], it is different from the one obtained by simply taking the complex logarithm of the wave function. The quantization by means of the QHJE is more involved than by means of the SE, but brings to the same results and moreover allows to investigate the classical limit. This is not possible with the usual approach, because the SE loses its significance for $\hbar \to 0$, while instead the QHJE simply becomes the classical HJ equation. This feature makes possible to investigate how the quantum action and its gradient generate the classical action and momentum, respectively.

The boundary values needed now are the real parts of quantum actions $X_k(x)$ for each arc (the explicit dependence by the energy will often be understood), and their first derivatives $X'_k(x)$, which are connected with the amplitude of the arcs' wave functions. These quantities are computed by solving the 1-dim QHJE on each caustic's arc. The procedure is described in detail in [1], and will be very briefly summarised here.

Firstly, the wave functions on the caustic's arcs are looked for in the exponential form:

$$\psi_k(x) = \exp(\frac{i}{\hbar} W_{Q,k}(x)) \qquad (17),$$

where the complex phase $W_{Q,k}(x)$ is the quantum action on the k-th arc, and is written as:

$$W_{Q,k}(x) = X_k(x) + i\, Y_k(x) \qquad (18).$$

In the limit $\hbar \to 0$ the imaginary part $Y_k(x)$ vanishes, and the real part $X_k(x)$ becomes the classical action. By inserting the Eqs. (17) and (18) into Eq. (3) and separating the real and imaginary parts, one gets the couple of 1-dim equations:

$$-X_k'^2(x) + Y_k'^2(x) + \frac{Y_k'^2(x) g_k'(x)}{g_k(x)} - Y_k''(x) = \left(\frac{2m}{\hbar^2}\right)(E - U_k(x)) g_k^2(x) \qquad (19),$$

and

$$-2 X_k'(x) Y_k'(x) - \frac{X_k'(x) g_k'(x)}{g_k(x)} + X_k''(x) = 0 \qquad (20).$$

The solution of the last equation is:

$$Y_k(x) = \hbar \, \mathrm{Log} \sqrt{\frac{X_k'(x)}{g_k(x)}} \qquad (21),$$

and by using this equation, Eq. (19) becomes:

$$+3\hbar^2 g_k'^2(x) X_k'^2(x) + 4 g_k^2(x) X_k'^4(x) - 3\hbar g_k^2(x) X_k''^2(x) +$$
$$2\hbar^2 g_k^2(x) X_k'(x) X_k'''(x) - 2\hbar^2 g_k(x) g_k''(x) X_k'^2(x) = 8m(E - U_k(x)) g_k^4(x) X_k'^2(x)$$

$$(22).$$

When $\hbar \to 0$, Eq. (22) turns into the CHJE on the k-th arc for $X_k(x)$, therefore this latter function, i.e. the real part of the quantum action, in this limit becomes the classical action $X_{\mathrm{cl},k}(x)$. Analogously, its derivative generates the classical momentum $p_{\mathrm{cl},k}(x)$. Outside the arc, along its continuation, the quantum action has only the imaginary part $Y_k(x)$, which satisfies the same equation (19), but with $X_k(x) = 0$. When a solution $X_k(x)$ of Eq. (22) is known, Eq. (21) gives $Y_k(x)$, and from Eqs. (18) and (17) one gets a complex solution of the SE on the arc. Finally, by combining this latter and its complex conjugate, the wave function on the arc is written in the trigonometric WKB-like form [7]:

$$\psi_k(x) = \frac{c_k}{\sqrt{\frac{X'_k(x)}{g_k(x)}}} Sin\left(\frac{X_k(x)}{\hbar} + \frac{\pi}{4}\right) \qquad (23),$$

where $c_k$ is a constant. Eq. (23) is an exact representation of the wave function on the caustic's arc, similar to the approximate WKB one, but differently from this latter, it holds at the turning points too, i.e. at the arc's ends, where instead the WKB amplitude diverges. Beyond the arc's ends, the wave function has another exact representation, of exponential form, depending only by the imaginary part $Y_k(x)$ of the quantum action.

Also in this case, a tentative value for the energy E and one initial point on the equipotential line are chosen, so to get the corresponding caustic. Then one has to solve the 1-dim QHJE (22) for $X_k(x)$ on each caustics' arc, construct the related arc wave functions from Eq. (23) and verify if along the caustics' arcs the quantization conditions are satisfied. These are discussed in [1, 7], and as for the case of the SE discussed in Sect. 2, require that each caustic' arcs can host the full oscillating part of a 1-dim wave function $\psi_k(x)$, with the same energy E and with its own potential energy $U_k(x)$. After found the energy eigenvalue, the related caustic and the arcs' wave functions, the amplitudes of these latter are, according to Eq. (23):

$$A_k(x) = \frac{c_k}{\sqrt{\frac{X'_k(x)}{g_k(x)}}} \qquad (24),$$

where the constants $c_k$ are chosen so that the four arc's wave functions have the same value at the common ends of two adjacent arcs. These amplitudes $A_k(x)$, together with the actions $X_k(x)$, give the boundary values for the exact system of equations (10) and (11, which can now be numerically solved both inside and outside the caustic. As the caustic can be oriented in two ways, two quantum actions and two partial wave functions are obtained, as discussed for the WKB case. The full wave function is their x-symmetrized or anti-symmetrized combination, so to respect the x-symmetry of the Barbanis Hamiltonian. In this way, both the trajectories' sub-families with the same caustic contribute to the full wave function.

Some results are presented in the Figs. 11-16. For simplicity, only the solutions inside the caustic are plotted.

The Fig. 11 contains the 3-dim graph of the real part $X_{v1}(x, y)$ of the quantum action $W_{Q,v1}(x, y)$ and its contour plot, for the state (2, 2) of the Barbanis system. The caustic was oriented starting from the left lower vertex v1. The energy computed for this state by means of the QHJE is $E_{HJ}$ = 5.18871, slightly different from the corresponding value from the SE approach, which is $E_{SE}$ = 5.18266. The comparison with the corresponding classical quantities in Fig. 6 shows that the real part of the quantum action follows waving the corresponding classical quantity, as it happens in the 1-dim case too [7].

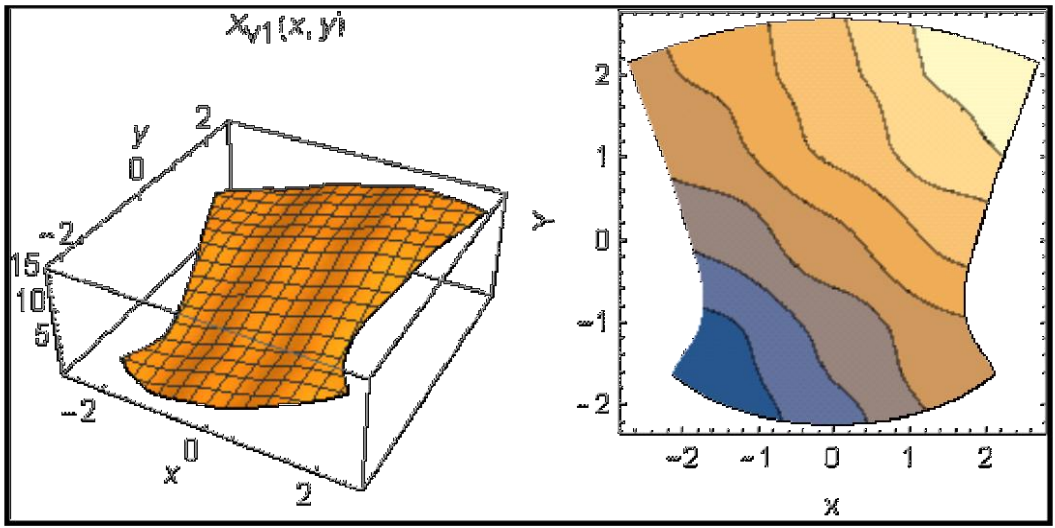

*Fig.11. The 3-dim plot of the real part $X_{v1}(x, y)$ of the quantum action $W_{Q,v1}(x,y)$ together with its contour plot, for the state (2, 2) of the Barbanis system. These quantities are computed with the caustic oriented starting from the left lower vertex v1. The energy computed for this state by means of the QHJE is $E_{HJ}$ = 5.18871, to be compared with the one obtained from the SE, which is $E_{SE}$ = 5.18266.*

The Fig. 12 contains the plots of the quantum amplitude A(x, y) for the wave function (2, 2) of the Barbanis system. The amplitude does not depend on the choice of the initial vertex. Differently from the corresponding WKB amplitude, in the quantum case this function oscillates inside the classical region with peaks and valleys, and is finite also on the caustic, where the WKB amplitude diverges.

The Fig. 13 contains the plots for the first partial wave function $\psi_{HJ,v1}(x, y)$ obtained as a real function from the complex wave function in the Eq. (9). The caustic was oriented from the vertex v1, and the wave function was computed according to the equation:

$$\Psi_{HJ, v1}(x, y) = A(x, y) \sin\left(\frac{1}{\hbar} X_{v1}(x, y)\right) \quad (25).$$

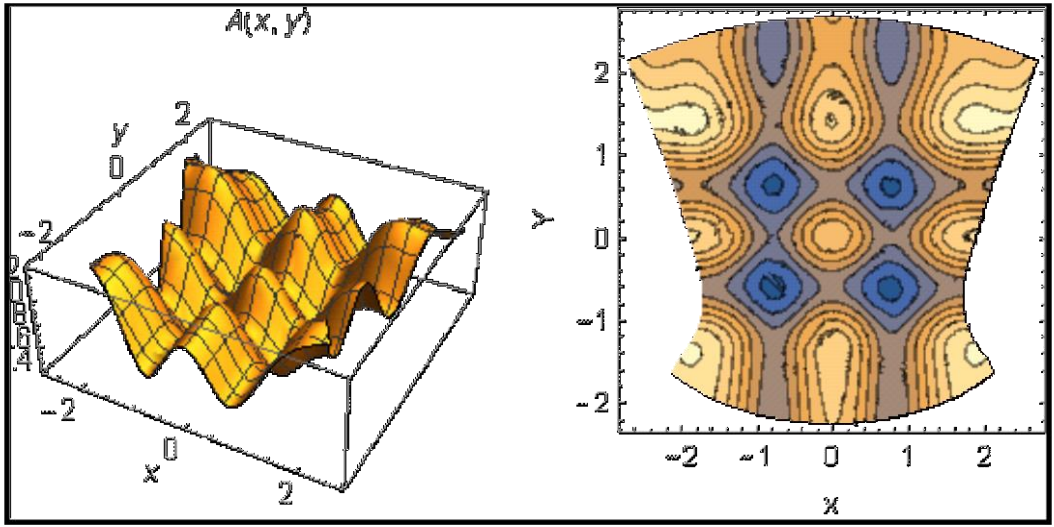

Fig.12. The plots of the quantum amplitude A(x, y) for the wave function (2, 2) of the Barbanis system, computed as expained in the text, by solving the boundary problem on the caustic for the exact QHJ equations (10) and (11). The amplitude does not depend on the initial vertex.

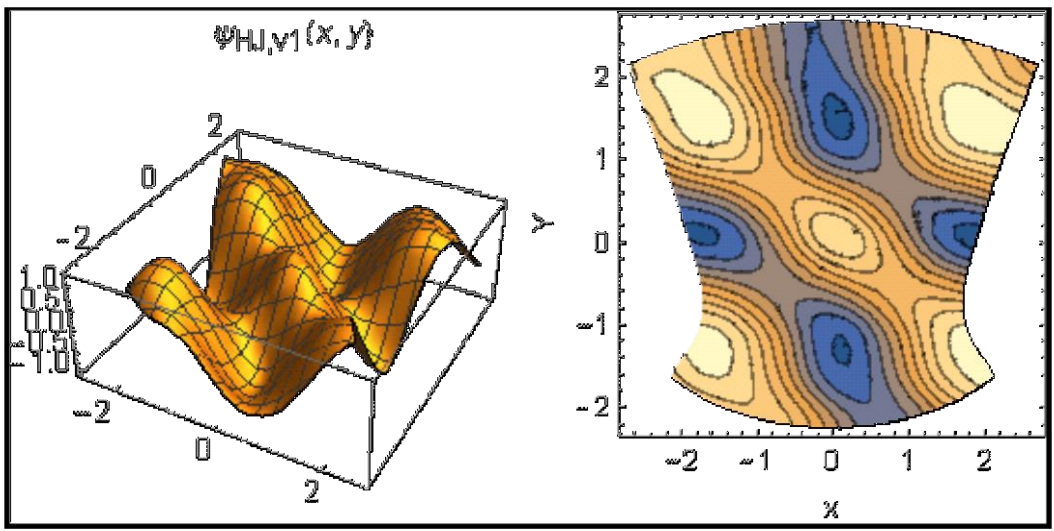

Fig. 13. The first partial wave function $\psi_{HJ,v1}(x, y)$ given by the imaginary part of the wave function in the Eq. (9), with quantities computed by orienting the caustic from the vertex v1.

The graphs for the second partial wave function, constructed by starting from the right lower vertex v2, are presented in the Figs. [14] and [15].

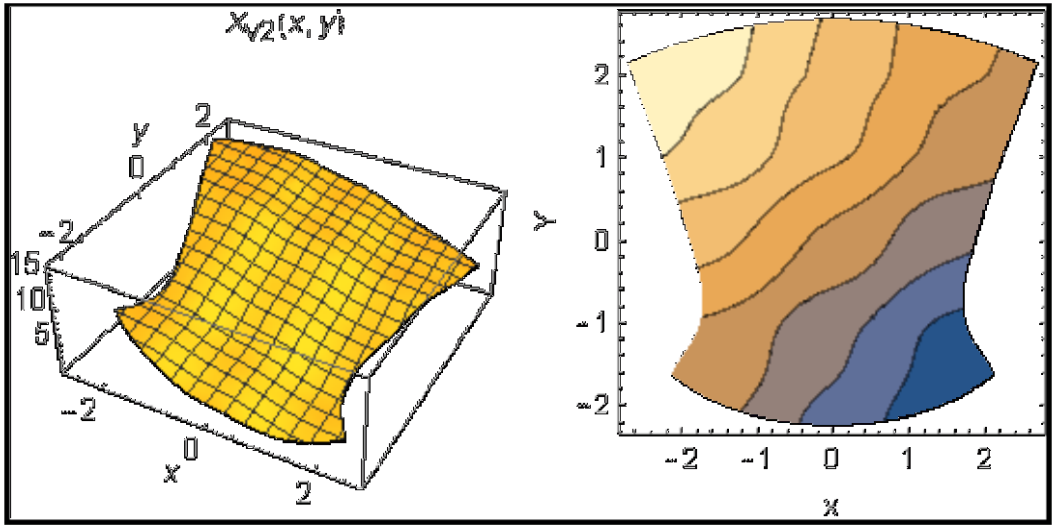

*Fig. 14. The 3-dim plot of the real part $X_{v2}(x, y)$ of the quantum action $W_{Q,v2}(x, y)$ together with its contours plot, for the state (2, 2) of the Barbanis system, computed by orienting the caustic starting from the right lower vertex v2.*

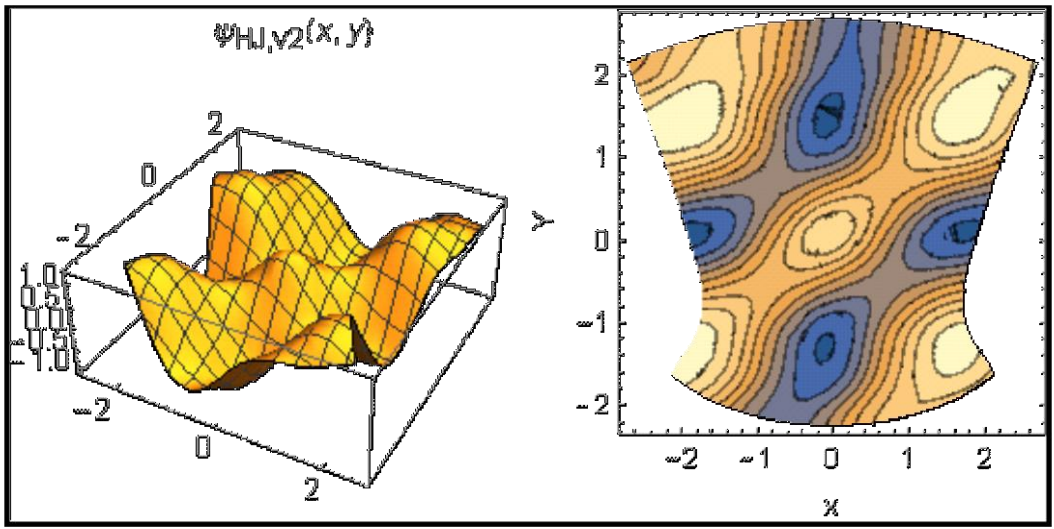

*Fig. 15. The second partial wave function $\psi_{HJ,v2}(x, y)$ computed by orienting the caustic from the right lower vertex v2.*

Finally, in Fig. 16 are the plots for the full x-symmetric wave function $\psi_{HJ}(x, y)$ for the state (2, 2). This is obtained by summing the two partial wave functions $\psi_{HJ,v1}(x, y)$ and $\psi_{HJ,v2}(x, y)$, computed by orienting the caustic from the vertices v1 and v2, respectively. The comparison between this wave function with the one in Fig. 3, shows a satisfactory, even if not exact, agreement with the wave function computed through the SE.

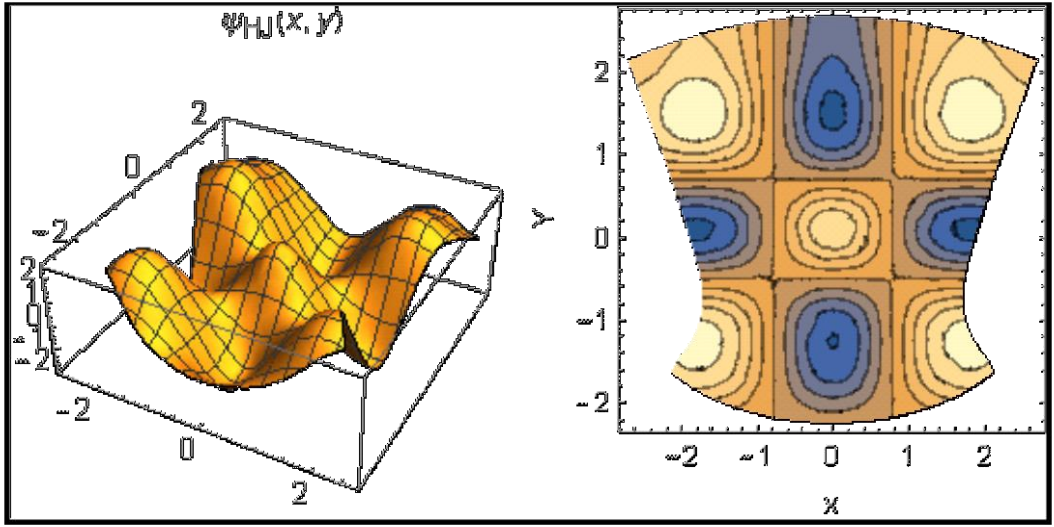

*Fig. 16. The plots for the full $\psi_{HJ}(x, y)$ wave function for the state (2, 2). This is obtained by x-symmetrization of the two partial wave functions $\psi_{HJ,v1}(x, y)$ and $\psi_{HJ,v2}(x, y)$, computed by orienting the caustic from the vertices v1 and v2, respectively.*

## 6. CONCLUSIONS

Both the presented approaches, the one based of the SE and that based on the QHJE, allow to compute the energy eigenvalues and to construct the quantum wave functions. The results presented in this paper demonstrate the correctness of the initial assumption, that each wave function for integrable non-separable Hamiltonian systems is one-to-one connected to a family of classical trajectories in the configuration space. These trajectories are the projections of the corresponding ones belonging to the invariant tori which foliate the phase space of an integrable system [8]. The results of the paper show that to each quantum wave function for these systems it univocally corresponds an invariant torus.

The approach by means of the SE is simpler and slightly more accurate than the QHJE one, but this is essentially due to the more involved procedure of this latter case and to the greater difficulty to numerically treat the nonlinear system of Eqs. (10) and (11), with respect to the single, linear SE. However, the QHJ approach is the only one which allows to completely investigate the links between classical and quantum mechanics. Indeed, while the SE loses its significance when h -> 0, this does not happens to the QHJE, which becomes the classical Hamilton-Jacobi equation. Traditionally, this latter equation is considered only as a mean to integrate the Hamilton's equations, through its complete integrals (when can be found), but this is a rather reductive vision. Actually, the special solutions of the CHJE are very important too, as describing the properties of families of classical trajectories and as limits of corresponding quantum actions. As shown in [7], only by means of exact (analytical or numerical) solutions of the QHJE it can be seen how the quantum quantities generate the corresponding classical ones, i.e. the classical action and the classical momentum. In this process, the complex phase of the wave function, i.e. the quantum action, has the main role, and displays its importance as fundamental dynamical quantity.

The link between the quantum and the classical mechanics only partially can be investigated by the way of the WKB approach, in which one tries to construct the quantum wave function by means of classical quantities. This is due to the increasing complexity of the WKB equations and the divergences of its series expansions. An effective procedure is the inverse, to construct the classical quantities from the quantum ones, computed from the QHJE.

Usually, the QHJE is derived from the SE, but this path can be inverted, and the SE can be equivalently derived from the QHJE. This latter equation can so be postulated and put as the basis of a formulation of the quantum mechanics, which includes both the traditional SE-based one, and the Hamilton-Jacobi formulation of the classical mechanics as its limit for h -> 0. This approach, more involved with respect to the SE, was not feasible at the birth of the quantum mechanics, but is conceptually more satisfying, and is now viable, due to the analytical advancements and the use of computers.